\documentclass[twocolumn]{cinc}
\usepackage{graphicx}
\usepackage[breaklinks]{hyperref}
\usepackage{breakurl}

\usepackage{makecell}

\usepackage{listings}
\usepackage{color}

\definecolor{dkgreen}{rgb}{0,0.6,0}
\definecolor{gray}{rgb}{0.5,0.5,0.5}
\definecolor{mauve}{rgb}{0.58,0,0.82}

\begin{document}
%\bibliographystyle{cinc}

% Keep the title short enough to fit on a single line if possible.
% Don't end it with a full stop (period).  Don't use ALL CAPS.
\title{{Building a Mastodon Compatible Java Server for ActivityPub}}

% Both authors and affiliations go in the \author{ ... } block.
% List initials and surnames of authors, no full stops (periods),
%  titles, or degrees.
% Don't use ALL CAPS, and don't use ``and'' before the name of the
%  last author.
% Leave an empty line between authors and affiliations.
% List affiliations, city, [state or province,] country only
%  (no street addresses or postcodes).
% If there are multiple affiliations, use superscript numerals to associate
%  each author with his or her affiliations, as in the example below.

\author {Sean Nian, Angela Huang, Ben Reed
\ \\ % leave an empty line between authors and affiliation
San José State University, San José, United States of America }

\maketitle
% LaTeX inserts the ``Abstract'' heading in the proper style and
% sets the text of the abstract in italics as required.
\begin{abstract}
ActivityPub is a decentralized social networking protocol that has gained significant attention from the media for its ability to communicate through the Fediverse, short for the federated web.
Servers such as Mastodon implement the ActivityPub protocol to communicate over the Fediverse.
In this paper, we deconstruct the core protocols used to build the distributed servers of the Fediverse.
We explore Mastodon's complex implementation of ActivityPub and created our own Mastodon instance using Java Spring Boot and ActivityPub to interoperate with Mastodon servers. 

% Of course, you must insert blank lines
% between the paragraphs of your LaTeX input file, since this is the
% only way to indicate paragraph boundaries.  Make sure that no blank
% lines appear between paragraphs in the formatted output, however.)
% 
% 
\end{abstract}
% LaTeX inserts the extra space here automatically.

\section{Introduction}
% Section numbering is automatic.  The examples on the next page
% illustrate how to make subsections.

Social media platforms such as Twitter (now X) and Facebook are used by hundreds of millions of people across the world.
However, because these platforms are owned by corporations that have full control, user data and privacy are vulnerable to misuse.
In addition, the central authorities can easily censor content.

In October 2022, after Elon Musk took over Twitter, users started searching for Twitter alternatives. One popular solution is to use decentralized social media networks.

Decentralized networks have no central authority, which means that they can be user-hosted.
Individuals who operate their servers have full control over their data and privacy, as well as control over the content that is posted on their servers.
Despite the decentralization of servers, users are still able to interact seamlessly with users on other servers.
Additionally, due to the decentralized nature of the platforms, users can keep their information such as followers and following lists if transferred to other servers.

\begin{figure}
\centering
\includegraphics[width=.8\linewidth]{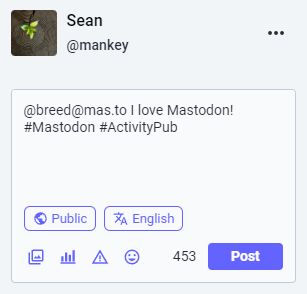}
\caption{Example Mastodon post.}
\label{fig:exampleMastodon}
\end{figure}

The most popular decentralized network that users flocked to after the Twitter takeover was Mastodon ~\cite{mastdoc}, which over a million users signed up for in early November 2022~\cite{twittermigration}. 
As shown in Figure~\ref{fig:exampleMastodon}, Mastodon is a microblogging platform that allows users to create short posts, called "toots," and contains many similar features to Twitter.

Mastodon is part of the Fediverse, short for federated universe, which is a network of interconnected decentralized social platforms. 
In this network, users can communicate with each other using a protocol called ActivityPub regardless of the software they are running~\cite{mastodonchallenges}.
As seen in Table 1, there are many different services offered by the Fediverse, ranging from social media type platforms to practical uses such as publishing, event planning, and more.
Other companies such as Meta are interested in joining the Fediverse, as they see the benefits of interoperability between services and improved accessibility of content~\cite{Meta}.

\begin{table}
    \centering
    \begin{tabular}{|c|c|c|c|c|}
    \hline
         Microblogging & Multimedia & Publishing  & Aggregator \\ \hline
         \makecell{Mastodon \\ (7.3m)} &  \makecell{PeerTube \\ (321k)} &  \makecell{Writefreely \\ (120k)} & \makecell{Lemmy \\ (417k)}  \\ \hline
         \makecell{Misskey \\ (744k)} &  \makecell{Pixelfed \\ (216k)} &  \makecell{BookWyrm \\ (26k)} &  \makecell{Kbin \\ (62k)}\\ \hline
         \makecell{Pleroma \\ (140k)} & \makecell{Funkwhale \\ (10k)} & \makecell{WordPress \\ (21k)} &  \makecell{Lotide \\ (283)}\\ \hline
    \end{tabular}
    \caption{Internet Services that use ActivityPub. \protect\cite{fedidb}}
    \label{tab:my_label}
\end{table}

%maybe explain how mastodon enables instances 
We have a project called Moth (short for Mammoth) that is compatible with Mastodon's use of ActivityPub and Mastodon's client API. 
As Mastodon servers are written in Ruby, a language that most students are not familiar with, we hope to create a more familiar environment for future research projects.
In this paper, we:
\begin{itemize}
    \item \textbf{Decontruct the messaging protocol of ActivityPub:} Explain how ActivityPub achieves server-to-server interoperability through the Fediverse.
    \item \textbf{Study Mastodon's implementation of ActivityPub:} Explain how Mastodon integrates ActivityPub in their services to communicate in the Fediverse.
    \item \textbf{Create our own Mastodon instance:} Explain how sending and viewing messages through the Fediverse is not as easy as one may think. 
\end{itemize}

\section{Background}

We use three key protocols in Moth: ActivityPub, Mastodon, and WebFinger. 
Mastodon is a self-hosted decentralized network platform that belongs to the Fediverse. 
ActivityPub is the most popular networking protocol that is used by the Fediverse and is used by Mastodon as well. 
Webfinger is a discovery protocol that Mastodon uses to retrieve information about users on other platforms within the Fediverse.

\subsection{ActivityPub}
\label{sec:activitypub}
ActivityPub~\cite{activitypub} is a protocol standard established by the World Wide Web Consortium (W3C). 
ActivityPub provides two REST API protocols: a server-to-server federation API and a client-to-server API. 
When implementing the ActivityPub protocol, servers must provide HTTPS URI endpoints that allow various actions such as creating posts, viewing messages, and interacting with users.
User endpoints are also defined in ActivityPub, where users are assigned unique URIs such as their personal inbox and outbox, which users may utilize to send and receive messages. 

\begin{figure}
\centering
\includegraphics[width=1\linewidth]{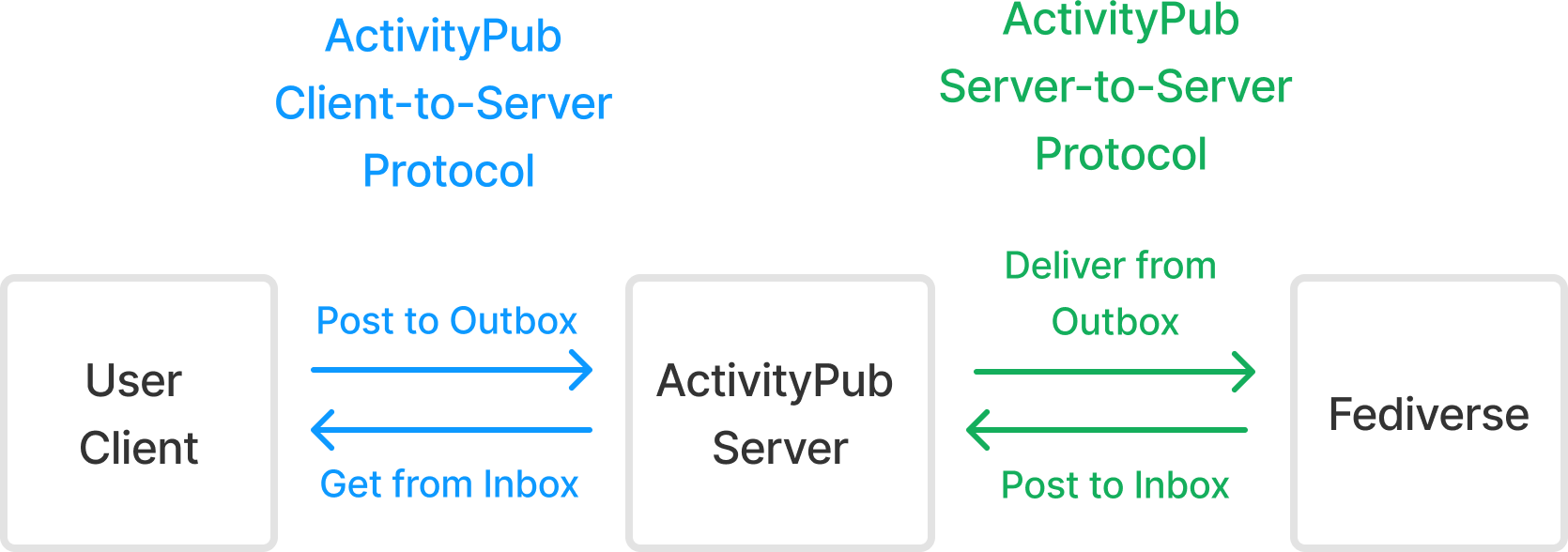}
\caption{Illustrates how the two types of ActivityPub protocols are used to enable communication between the Fediverse, ActivityPub servers, and user clients.}
\label{fig:activityPubProtocols}
\end{figure}

Platforms that implement the server-to-server API are able to share information with each other, allowing users on different platforms to interact. 
On the other hand, the client-to-server API enables users to communicate with ActivityPub. 
Figure~\ref{fig:activityPubProtocols} shows how ActivityPub protocols enable communication between many platforms in the Fediverse. 

\subsection{Mastodon} 
\label{sec:mastodon}
Mastodon, a social media platform that belongs to the Fediverse, is a microblogging service that functions similarly to Twitter. 
Mastodon client applications (on mobile phones or the web) use Mastodon-specific client REST APIs.
Along with most of the Fediverse, it implements the ActivityPub protocol. 

Rather than a centralized server, Mastodon consists of a network of servers called instances.
Each instance is independently run, but they are able to communicate with one another by implementing the ActivityPub protocol. 
Communication between members in the same instance is done internally in the instance.

When delivering messages to users in different servers across the Fediverse, ActivityPub's protocol is used.
The remote users' HTTPS URIs must first be found in order to locate their inbox.
To achieve this, the WebFinger protocol must be used~\cite{mastodonwebfinger}.

A not-well-documented portion of Mastodon is its representation of users and which servers they are hosted on.
Mastodon uses an @username@server notation when sending server-to-server messages to other Mastodon instances.
The same username can be used across different instances of Mastodon, so the @username@domain syntax is used to determine which instance the user is from. 
@domain is the name of the Mastodon instance, for example, mastodon.social, a popular Mastodon instance. 

\subsection{WebFinger} 
\label{sec:webfinger}
When transmitting messages to different domains and services, the @username@server mention format is not enough to find a remote user's inbox and outbox.
Therefore, mentions must first be sent to an HTTPS URI in order to find these links. 

WebFinger is a tool that can resolve the links to a resource, given the URI. 
After a user sends an HTTP WebFinger request with the user mention, WebFinger will return a JSON object that contains information about the user. 
This object can include a user's links, preferred avatar, and identity service ~\cite{ietfwebfinger}. 
Along with the user URI, which allows remote Actors' inboxes and outboxes to be found, the user's public key is also shared, which can be used to establish a secure connection.

\section{ActivityPub Vocabulary}
All ActivityPub servers must share these common specifications.
Without these common traits, an ActivityPub server would not be compatible with the Fediverse. 

\subsection{Objects}
We need to represent posts and people. 
We do this with ActivityPub objects.
By convention, ActivityPub objects are serialized using JSON format, inheriting the advantages JSON has, such as parsability and human readability~\cite{JSON}.
Two notable ActivityPub objects that will be referred to often are: 

\begin{itemize}
    \item \textbf{Actor:} An Actor may represent a type of application, group, organization, person, or service. 
    \item \textbf{Note:} A Note is an object with an attached content String.
\end{itemize}

Combining these terms, ActivityPub's underlying idea is simple: Actors send and receive Notes. 

\subsection{Activities}
In the ActivityPub protocol, events are represented by Activities, which allow users to interact within the Fediverse. 
Activities define a wide variety of actions, such as "Create" a message, "Follow" a user, or "Like" a post to name a few.
ActivityPub activities are also serialized using JSON format. 

Activities and Objects in ActivityPub are highly customizable between servers of the Fediverse.
Depending on the type, different Activities may include different fields.
A "Create" Activity may include the Object it is attempting to create, such as a "Note".
On the other hand, a Follow Activity may address the user being followed. 
A server may also choose to include different metadata in their activities.
The paper will later address how Mastodon implements their version. 
However, at the bare minimum, each server must produce Activities that include the type of action taken ("Create", "Follow", "Like"), as well as the user performing the activity, represented as an "Actor" object. 
    
\subsection{Endpoints}
Communication through ActivityPub occurs when an Activity is sent or retrieved from an API endpoint.
An endpoint is a URL where ActivityPub servers and users can utilize HTTP requests such as POST and GET to interact with a server or another user. 
In ActivityPub's server-to-server communication, each server must have its own set of endpoints; however, its functionality and response format are highly customizable in an individual server. 

Moreover, it is important to note that a server's endpoints have different functionality than a user's endpoints. 
A server's endpoints typically handle server-to-server communication, while a user's endpoints handle client-to-server interactions.
Some notable ActivityPub endpoints are:

\begin{itemize}
    \item \textbf{Outbox:} A user may utilize a POST request to place Activities into their outbox endpoint. 
    Subsequently, a server can employ a GET request to retrieve Activities from a user's outbox endpoint. 
    \item \textbf{Inbox:} A user may utilize a GET request to retrieve Activities from their inbox endpoint. 
    Subsequently, a server can employ a POST request to send Activities to a user's inbox endpoint. 
    \item \textbf{Followers, Following, Likes, Shares:} A user may have these endpoints to denote who follows them, who they follow, what messages they like, and what messages they have shared. 
    Information from this endpoint may also be publicly accessed by utilizing HTTP GET requests on the URI.
\end{itemize}

A server does the heavy lifting in server-to-server communication. 
Once a user posts to their outbox, it is the server's responsibility to GET its user's activities from their outbox and POST it to the desired inbox. 
Otherwise, federation in the Fediverse would not be effective.

\section {Mastodon's Implementation}
Mastodon defines a custom client API that is compatible with the ActivityPub protocol. 
Each Mastodon instance must implement both Mastodon's client API and the ActivityPub protocol to be supported on Mastodon client applications (on mobile phones or the web) and to be able to communicate server-to-server in the Fediverse. 

\begin{figure}
\centering
\includegraphics[width=1\linewidth]{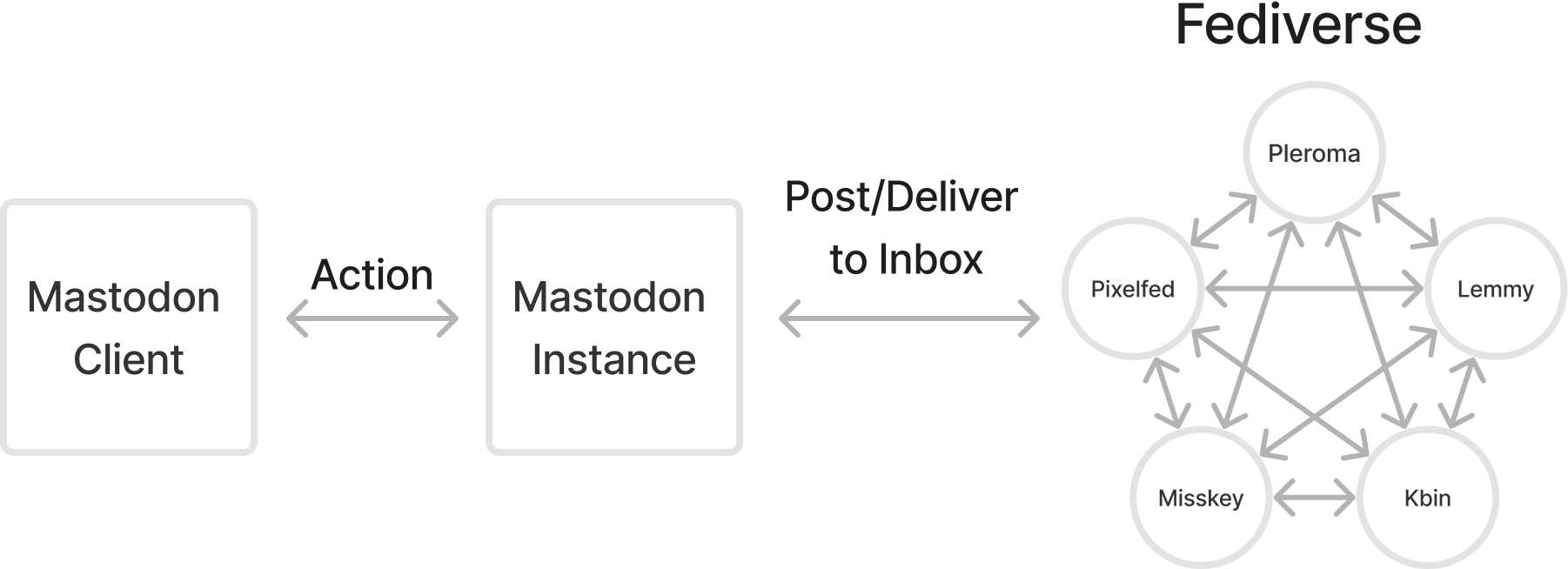}
\caption{Depicts user interaction with the Mastodon Client and how the Mastodon instance interacts with the rest of the Fediverse with the ActivityPub protocol.}
\label{fig:mastodonProtocols}
\end{figure}

Mastodon client API for client-to-server interaction makes user management, message displaying, and discovering messages across the Fediverse much easier. 
Client-to-server interaction is streamlined so that users may interact with the Mastodon instance without needing to POST messages through the user's outbox and receive messages through the inbox. 
Additionally, Mastodon does not utilize ActivityPub's outbox like other ActivityPub servers.
Mastodon instances only send HTTP POST requests to other servers' inboxes and receive requests in their own inbox.
Figure~\ref{fig:mastodonProtocols} depicts the simplification that the Mastodon client provides for users. 
The next sections highlight Mastodon-specific design choices of incorporating server-to-server messaging through ActivityPub.

\subsection{Objects}
Mastodon also needs to represent posts and people.
Luckily, objects in Mastodon are designed to be similar to ActivityPub, allowing for near 1:1 mapping. 
On top of ActivityPub information, Mastodon's client API includes additional information for each object to support Mastodon-specific features.

\begin{figure}
\centering
\includegraphics[width=.8\linewidth,height=6.3cm]{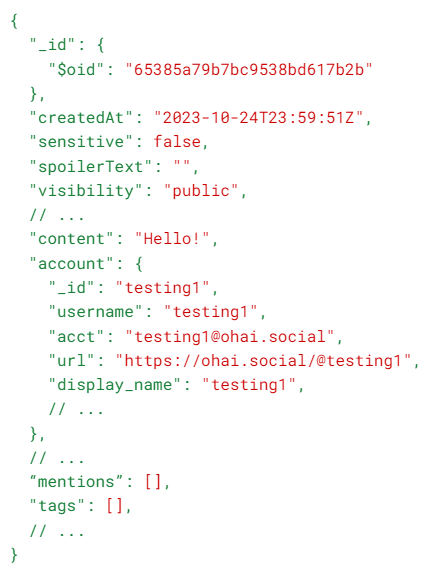}
\caption{Mastodon Status serialized in JSON converted from a received Note}
\label{fig:exampleMastodonStatus}
\end{figure}

Users in Mastodon are represented as Accounts, which are converted from ActivityPub Actors.
Mastodon builds on top of Actors by adding additional information such as a display name as shown in Figure~\ref{fig:exampleMastodonStatus}. 

As a microblogging service, Mastodon introduces Statuses to represent posts, which are similar to tweets on Twitter. 
Users on Mastodon may interact with Statuses by liking, boosting (reposting), favoriting, and bookmarking. 

In addition, Mastodon includes supplementary features that are built on top of the ActivityPub Note.
Figure~\ref{fig:exampleMastodonStatus} shows additional metadata that is included in Mastodon's implementation of messages.
Features such as visibility, mentions, and tags are special features that are non-normative in ActivityPub's protocol. 

\subsection{Activities}

\begin{figure}
\centering
\includegraphics[width=.9\linewidth]{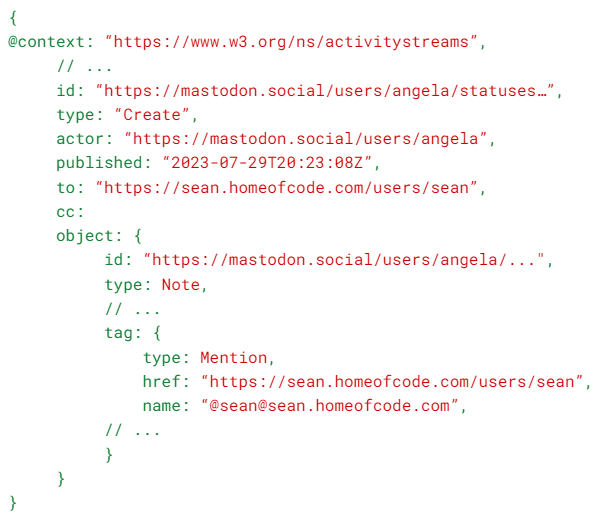}
\caption{A Mastodon converted Activity serialized in JSON received from another Mastodon instance}
\label{fig:exampleActivity}
\end{figure}

Since Activities are ActivityPub defined, Mastodon only builds on top of Activities; however, modifications are needed between Mastodon-defined objects and the rest of the Fediverse. 
As seen in Figure~\ref{fig:exampleActivity}, Mastodon's Activities contain information such as who it is directed toward, the object being sent (a Status converted into a Note), and the Actor who initiated the Activity.

Since Mastodon objects include extra information that may not be supported in other servers of the Fediverse, each Mastodon instance must convert Mastodon objects into ActivityPub objects. 
A Mastodon instance must convert Actors into Accounts when they receive messages from the Fediverse, and convert Accounts into Actors when sending Activities to other servers. 
Additionally, each instance must convert ActivityPub Notes into Mastodon Statuses to display them on Mastodon's client interface and convert Statuses into Notes when attaching them to Activities.
This way, Fediverse servers that implement ActivityPub are compatible with sending and receiving messages to all Mastodon instances.  

\subsection{Mentions}
As seen in Figure~\ref{fig:exampleMastodonStatus}, a Status may include mentions, which are stored in the Status itself. 
When a user is mentioned in a Status body, Mastodon will automatically forward that Status to the mentioned user's server.
This is a non-normative feature of ActivityPub, where "@username" will often route messages toward a user's inbox. 

%reword tags
\subsection{Tags}
Finding messages across the Fediverse may be difficult for users, as there is no way to search for specific messages in ActivityPub's protocol.
Mastodon fixes this problem by introducing tags, more commonly known as hashtags. 
As shown in Figure~\ref{fig:exampleMastodonStatus}, tags may be included in a Status body.
A Mastodon server should group Statuses with the same tag into a collection of Statuses.
This enables the discovery of messages on Mastodon as users can find similar Statuses of their topic of interest by sorting Statuses by tags. 

%visibility in mastodon 
\subsection{Visibility}
In Figure~\ref{fig:exampleActivity}, the "to" property designates the primary Actor that the Activity is directed toward.
However, the visibility of Statuses in Mastodon is designated by the visibility field in Figure~\ref{fig:exampleMastodonStatus}. 

The visibility field in Mastodon Statuses handles who has permission to see a post. 
Instead of a client-to-server interaction where a user would GET messages from their inbox and POST messages to their outbox, a Mastodon instance simply stores a Status in their database with the visibility set.
This way, the server would not have to send an individual Activity to every user's inbox. 

The visibility field also enables the implementation of private messages in the ActivityPub protocol. 
To send private messages in Mastodon, a Status is sent with the visibility set to private.
A private visibility means that only users who have been tagged in the Status can view the Status.
This also enables group conversations within Mastodon, as multiple users may be tagged in a Status. 

\subsection{Following an Account}
Mastodon users may follow users from other Fediverse servers that implement ActivityPub.
When a Mastodon user follows an Account from another Mastodon instance, the followed Account's Statuses should be automatically forwarded to the followee's host server.

\subsection{Account Deletion}
When an Account is deleted from a Mastodon instance, a global Activity should be sent out to all Mastodon instances.
The Activity will have a "Delete" type and will include the Account being deleted.
This is an attempt to synchronize databases through all Mastodon instances, ensuring that no server wastes storage space to keep a user that has been deleted.
Servers then may choose to delete existing information from the deleted Account. 

\section{Results}
In our project, we created our own Mastodon instance compatible with Mastodon's client API and ActivityPub called Moth using Java Spring Boot and MongoDB.
Java Spring Boot supports JSON natively and makes implementation easier with magical Spring Boot mapping.

We created ActivityPub Actors from Mastodon Accounts for server-to-server communication in order to send messages through the Fediverse.
We defined simple Java classes with fields and types matching the JSON object types for both ActivityPub and Mastodon.
SpringBoot automatically serialized these types to and from HTTP requests and MongoDB operations.

Two major difficulties we faced were 1) mapping field names between ActivityPub and Mastodon and 2) understanding which fields were required and which were nullable.
A compounding difficulty was the lack of errors back: if an ActivityPub server sent a request and was not happy with the answer, often we would get nothing back, so debugging was challenging.

\section{Related Work}
There are many related platforms within the Fediverse, as almost all of the platforms implement ActivityPub. Some of these include Misskey~\cite{misskey} (microblogging), Lemmy~\cite{lemmy} (link aggregator), and PixelFed~\cite{pixelfed} (photo sharing). Users with accounts on Mastodon are able to interact with users of Misskey, Lemmy, PixelFed, and other platforms that implement ActivityPub.

\section{Conclusion}
By deconstructing the architecture behind ActivityPub and Mastodon, we gain a deeper understanding of decentralized social media networks.
The source code for our implementation of a Mastodon instance can be found on SJSU CS Systems Group's GitHub~\cite{moth}.

\balance

%\section*{Acknowledgments}  
% This section is not numbered.
%
% Thank you Meta for funding.
%Give any acknowledgments here.

% LateX generates the ``References'' heading automatically and switches
% to 9 point type for the bibliography.  Please  use BibTeX and
% follow the examples in the sample 'refs.bib' file to enter your references.
%\bibliography{refs}

% If you don't use BibTeX (why not?) , comment out or remove the previous
% line, and uncomment the following lines up to the ``}\end{bibliography}''
% line below:
%\begin{thebibliography}{99}{ %\small
% \bibitem{tag} (General form) J. K. Author, ``Name of paper,''
%   \emph{Abbrev. Title of
%   Periodical}, vol. x, no. x, pp. xxx--xxx, Abbrev. Month, year. 

% \bibitem{ito}  M. Ito et al., ``Application of amorphous oxide TFT to
%   electrophoretic display,'' \emph{J. Non-Cryst. Solids}, vol. 354, no. 19,
%   pp. 2777--2782, Feb. 2008.
  
% \bibitem{fardel}  R. Fardel, M. Nagel, F. Nuesch, T. Lippert, and
%   A. Wokaun, ``Fabrication of organic light emitting diode pixels by
%   laser-assisted forward transfer,'' \emph{Appl. Phys. Lett.}, vol. 91,
%   no. 6, Aug. 2007, Art. no. 061103.
  
% \bibitem{buncombe} J. U. Buncombe, ``Infrared navigation Part I: Theory,''
%     \emph{IEEE Trans. Aerosp. Electron. Syst.}, vol. AES-4, no. 3,
%     pp. 352--377, Sep. 1944.
      
% Uncomment the following line if you are not using BibTeX.
%}\end{thebibliography}

% LaTeX inserts the ``Address for correspondence'' heading.
%\begin{correspondence}
%Sean Nian\\
%1 Washington Sq, San Jose, CA 95192\\
%sean.nian@sjsu.edu
%\end{correspondence}

\end{document}